\documentclass[aps,preprint,10pt,superscriptaddress,footinbib]{revtex4-1}

\usepackage{amsmath}
\usepackage{amssymb}
\usepackage{amsfonts}

\usepackage{graphicx}

\usepackage{amsthm}

\newtheorem*{theorem*}     {Theorem}

\newtheorem*{lemma*}       {Lemma}

\newtheorem*{corollary*}   {Corollary}

\theoremstyle{remark}
\newtheorem*{remark*}      {Remark}

\newtheorem*{example*}     {Example}

\DeclareMathOperator{\Ad}{Ad}

\DeclareMathOperator{\Orth}{{O}}
\DeclareMathOperator{\SO}{{SO}}

\DeclareMathOperator{\U}{{U}}

\newcommand{\norm}[1]{{\left\| #1 \right\|}}
\newcommand{\spp}[1]{^{(#1)}}
\newcommand{\psp}{\spp}
\newcommand{\sbp}[1]{_{(#1)}}

\newcommand{\paren}[1]{{\left( #1 \right)}}
\newcommand{\normalparen}[1]{{( #1 )}}
\newcommand{\bigparen}[1]{\bigl({#1}\bigr)}

\newcommand{\normalbrac}[1]{{\{ #1 \}}}
\newcommand{\brac}[1]{{\left\{ #1 \right\}}}

\newcommand{\angl}[1]{{\left\langle #1 \right\rangle}}

\newcommand{\pder}[2]{{\partial{#1}\over\partial{#2}}}

\newcommand{\oder}[2]{{{d{#1}}\over{d{#2}}}}

\newcommand{\diag}{{\mbox{\rm diag}}}

\makeatletter
\def\loweq@align#1#2{\lower.6ex\vbox{\baselineskip\z@skip\lineskip\z@
    \ialign{$\m@th#1\hfil##\hfil$\crcr#2\crcr=\crcr}}}
\def\lowsim@align#1#2{\lower.6ex\vbox{\baselineskip\z@skip\lineskip\z@
    \ialign{$\m@th#1\hfil##\hfil$\crcr#2\crcr\sim\crcr}}}
\def\geqq{\mathrel{\mathpalette\loweq@align >}}
\def\leqq{\mathrel{\mathpalette\loweq@align <}}
\def\grsim{\mathrel{\mathpalette\lowsim@align >}}
\def\lesssim{\mathrel{\mathpalette\lowsim@align <}}
\def\gsim{\mathrel{\mathpalette\lowsim@align >}}
\def\lsim{\mathrel{\mathpalette\lowsim@align <}}
\def\lddots{\mathinner{\mkern1mu
    \raise\p@\hbox{.}\mkern2mu\raise4\p@\hbox{.}\mkern2mu
    \raise7\p@\vbox{\kern7\p@\hbox{.}}\mkern1mu}}
\makeatother
\newcommand{\grless} 
{ {\, \raise-.24em\hbox{$<$} \hspace{-0.8em} \raise.31em\hbox{$>$}\, } }
\newcommand{\lessgr} 
{ {\, \raise-.24em\hbox{$>$} \hspace{-0.8em} \raise.31em\hbox{$<$}\, } }

\newfont{\bg}{cmr10 scaled\magstep4}                    
\newcommand{\bigzerou}{\smash{\lower1.7ex\hbox{\bg 0}}}

\renewcommand{\a}{\alpha}
\renewcommand{\b}{\beta}

\renewcommand{\l}{\lambda}

\newcommand{\om}{\omega}

\newcommand{\s}{\sigma}

\newcommand{\tht}{\theta}

\newcommand{\p}{\partial}

\newcommand{\f}{\frac}
\newcommand{\q}{\quad}
\newcommand{\qq}{\qquad}

\newcommand{\nn}{\nonumber \\ }

\newcommand{\LL}{{\cal L}}

\newcommand{\CC}{{\cal C}}

\newcommand{\crl}[1]{[-\infty,\infty]}

\newcommand{\Ref}[1]{(\ref{#1})}

\newcommand{\wt}{\widetilde}

\newcommand{\Pb}[2]{\normalbrac{{#1},{#2}}}
\newcommand{\sa}{selfadjoint}

\newcommand{\alg}[1]{{\mathfrak #1}}

\newcommand{\ads}{\mathrm{AdS}}

\newcommand{\sfc}{{\cal S}}

\newcommand{\Xx}{x}
\newcommand{\Xy}{y}
\newcommand{\Xz}{z}
\newcommand{\Xw}{w}
\newcommand{\Xs}{s}
\newcommand{\Xt}{t}

\newcommand{\Lxy}{J_{xy}}
\newcommand{\Lxz}{J_{xz}}
\newcommand{\Lxw}{J_{xw}}
\newcommand{\Lxs}{\widetilde{K}_{x}}
\newcommand{\Lxt}{K_{x}}
\newcommand{\Lyz}{J_{yz}}
\newcommand{\Lyw}{J_{yw}}
\newcommand{\Lys}{\widetilde{K}_{y}}
\newcommand{\Lyt}{K_{y}}
\newcommand{\Lzw}{J_{zw}}
\newcommand{\Lzs}{\widetilde{K}_{z}}
\newcommand{\Lzt}{K_{z}}
\newcommand{\Lws}{\widetilde{K}_{w}}
\newcommand{\Lwt}{K_{w}}
\newcommand{\Lst}{L}

\newcommand{\se}[1]{\emph{#1.---}}

\renewcommand{\ge}{\geqslant}
\renewcommand{\le}{\leqslant}
\newcommand{\hv}{homogeneity vector}

\begin{document}

\vspace*{-.5cm}
\hspace{13cm} \hbox{OCU-PHYS-471/AP-GR-142}

\title{
  Cohomogeneity-one-string integrability of spacetimes
}
\author{Yoshiyuki Morisawa}
\email{morisawa@sci.osaka-cu.ac.jp}
\affiliation{Advanced Mathematical Institute, 
  Osaka City University, 
  Osaka 558--8585 Japan}
\author{Soichi Hasegawa}
\affiliation{Department of Mathematics and Physics, 
  Graduate School of Science, Osaka City University, 
  Osaka 558--8585 Japan 
}
\author{Tatsuhiko Koike}
\email{koike@phys.keio.ac.jp}
\affiliation{Department of Physics and REC for NS, 
  Keio University, Yokohama 
  223--8522 Japan
}
\author{Hideki Ishihara}
\email{ishihara@sci.osaka-cu.ac.jp}
\affiliation{Department of Mathematics and Physics, 
  Graduate School of Science, Osaka City University, 
  Osaka 558--8585 Japan
}

\date{September 22, 2017}

\begin{abstract}
We present a framework for reducing all possible cohomogeneity-one 
strings, i.e., strings with geometrical symmetry, 
in a given spacetime to mechanical systems,
and for analyzing integrability of the systems.  
As applications, it is clarified whether
the systems of cohomogeneity-one strings in $\ads_5$, 
$\ads_5 \times S^5$, 
and $\ads_5 \times T^{p,q}$ 
are integrable or not.
This method may reveal a different type of hidden symmetry of
spacetimes.
\end{abstract}

\maketitle

\se{Introduction}
Dynamics of extended object such as string
draws much attention in fundamental physics,
condensed matter physics, 
and cosmology~\cite{Vilenkin:2000jqa}.
Classical solutions of string dynamics have been widely studied. 
A variety of stationary solutions
and 
time evolution of strings 
are investigated~\cite{BurdenSRS, FrolovEtAl, 
VLS, Ogawa:2008qn,
Igata:2009fd} 
in spacetimes 
such as flat, de~Sitter, anti de~Sitter, and black hole spacetimes.

One can treat these string solutions in a unified manner~\cite{
IshiharaKozaki,
Koike:2008fs, 
Kozaki:2009jj}
because their 
world sheets  
share 
certain symmetry with the spacetimes. 
Such strings are called cohomogeneity-one (C1) strings 
and specified by the symmetry 
(Killing vector field) shared with the spacetime. 
In that class, the 
string equation of motion, 
which is a  partial differential equation, 
is reduced to a 
system of ordinary differential equations. 
Namely, motion of the string reduces to motion of a
free particle on a 
projected spacetime, 
called the orbit space, endowed with a certain weighted metric. 
Classification of all types of C1 strings has been carried out in
four-dimensional Minkowski spacetime~\cite{IshiharaKozaki} 
and in 
five-dimensional anti-de Sitter spacetime~\cite{Koike:2008fs}. 
It has also 
been shown in the former 
that all types of C1 strings are integrable, i.e., solved up to quadrature~\cite{Kozaki:2009jj}.

In this Letter, 
we develop a general framework 
of probing spacetime \emph{symmetry} by 
C1 strings. 
It is well-known that motion of a free particle 
follows a geodesic in the spacetime and 
integrability of the motion 
reflects the symmetry of the spacetime. 
In four-dimensional Schwarzschild spacetime, 
the isometries thereon immediately 
guarantee 
geodesic integrability. 
By contrast, in Kerr spacetime, 
the isometry group alone does not 
guarantee geodesic integrability.  
However, 
geodesics are in fact integrable, 
which reflects ``hidden symmetry.'' 
Namely, geodesic 
integrability is implied by the existence of 
Carter's constant~\cite{Car68}, 
which follows from 
existence of 
a nontrivial Killing tensor field~\cite{WalPen70}. 
This shows that 
particle motion on a given spacetime can reveal its ``hidden
symmetry.''  
In a similar sense, 
string motion may reveal a different type of hidden symmetry. 
We introduce the concept of \emph{C1-string integrability} of a spacetime, 
which means that \emph{all} allowed C1 strings are 
(Liouville) integrable. 
This is of course a necessary condition for integrability of all
classical (bosonic) strings.

Recently,
dynamics of classical string was widely studied in the context of
AdS/CFT correspondence. 
While string motion is 
integrable in $AdS_5 \times S^5$~\cite{AdS5xS5}   
chaotic string dynamics is reported
in AdS soliton~\cite{AdS_soliton}
and in $AdS_5 \times T^{1,1}$~\cite{AdS5xT11}.
The obtained chaotic solutions are also C1 strings. 
We will 
apply our method to 
those spacetimes 
and examine C1-string integrability. 
We will see that 
the 
chaotic behavior in $\ads_5\times T^{1,1}$ is 
naturally understood as absence of 
necessary 
number of 
conserved quantity for C1-string integrability. 
We will also see that 
two geodesically integrable spacetimes which have the same isometry group 
can be distinguished by C1-string integrability. 

We will present our method in such a way that one can 
fully exploit the symmetry of the 
spacetime.

\se{Cohomogeneity-one strings} 
\label{general}
Let $(M,g)$ be a $d$-dimensional spacetime. 
Let $G$ be 
the isometry group of 
$(M,g)$ and $\alg g$ be its Lie algebra. 
An element $\xi$ of $\alg g$ 
is called a Killing vector field on $(M,g)$ 
and generates a one-parameter
family of isometries. 
The string that shares a 
such a one-parameter family 
with the spacetime
$(M,g)$ is called a \emph{cohomogeneity-one (C1)} string. 
More precisely, 
a C1 string has a world sheet 
which is foliated by one-dimensional 
orbits 
of some Killing vector field $\xi\in\alg g$. 
We will call $\xi$ the \emph{{\hv} field}. 

Once the spacetime $(M,g)$ is given, one can 
study 
all C1 strings by considering all {\hv} $\xi\in\alg g$. 
However, 
two string solutions, i.e. 
the world sheets $\sfc$ and $\sfc'$, 
are physically the same 
if the world sheet $\sfc$ is sent 
to $\sfc'$ by an isometry $\phi\in G$, i.e., 
$\sfc'=\phi(\sfc)$. 
We therefore 
classify the C1 strings by this equivalence class. 
In terms of 
{\hv} fields, 
$\xi$ and $\xi'$ are equivalent if there exists 
$\phi\in G$ such that 
$\xi'$ is proportional to $\phi_*\xi$. 
In a more algebraic terms, 
each physically different string
corresponds to 
an element of $\alg g$ up to 
equivalence by $\Ad_G$ and up to scalar multiplication, 
where $\Ad_G$ is the conjugation by $G$. 
The classification has been carried out for 
four-dimensional Minkowski spacetime~\cite{IshiharaKozaki} and 
five-dimensional anti-de Sitter spacetime~\cite{
Koike:2008fs}.

\se{Dynamics} 
We consider a string 
that is governed by the Nambu-Goto action 
\begin{align}
  S=\f{1}{2\pi \a'}
  \int d^2\s\sqrt{g_{ab}\pder{x^a}{\s^\a}\pder{x^a}{\s^\b} }, 
  \label{eq-string-action}
\end{align}
where the prefactor $\a'$ is the string tension. 
If the string is C1, one can carry out the integral of 
\Ref{eq-string-action} in the direction 
of 
$\xi$. 
This yields
\begin{align}
  S=\f{l}{2\pi \a'}
  \int d\l\sqrt{fh_{ab}\oder{x^a}{\l}\oder{x^a}{\l} }, 
  \label{eq-NG-C1}
\end{align}
where 
$
  h_{ab}:=g_{ab}-{\xi_a\xi_b}/f, 
$
with $
f:=g_{ab}\xi^a\xi^b, 
$
is the projection metric 
normal 
to $\xi$ and $l$ is a constant with
a dimension of length. 
We define the orbit space $O$ as 
the quotient of $M$ by the orbits of $\xi$. 
Now, one observes that 
the action \Ref{eq-NG-C1} is nothing but 
the action 
for a free 
particle in the spacetime $(O,fh)$. 
Thus, 
the original problem of 
finding 
world sheets of a C1 string 
has been reduced 
to that of finding geodesics on the orbit space $O$ 
with the \emph{weighted metric} $fh$. 
In the sequel, 
it is beneficial to use the 
Polyakov-type action,  
\begin{align}
  S=
  \f{l}{4\pi \a'}
  \int d\l\paren{\f1Nfh_{ab}\oder{x^b}{\l}\oder{x^a}{\l} +N}, 
  \label{eq-Poly-C1}
\end{align}
which 
is equivalent to \Ref{eq-NG-C1} and 
does not involve a square root. 
The equivalence can be seen 
by taking variations of 
\eqref{eq-Poly-C1} 
with respect to $N$  
and then inserting the resulting equation 
back to 
\eqref{eq-Poly-C1}. 
Note that the function $N(\l)$
appearing in the equation of motion (EOM) 
can be chosen arbitrarily 
because of the
parametrization invariance of the action. 
We shall lift he problem 
on $O$ again 
to that on $M$ so 
as to use the high symmetry of $(M,g)$. 
Now, consider an action defined on $(M,fg)$: 
\begin{align}
  S
  =
  \f{l}{4\pi \a'}
  \int d\l
  \paren{\f1N
    fg_{ab}\oder{x^a}{\l}\oder{x^b}{\l} 
    +N
  }. 
  \label{eq-lifted-action}
\end{align}
Then, a solution to the EOM derived from 
the action \eqref{eq-Poly-C1} 
is a solution to the EOM derived from 
\eqref{eq-lifted-action} which satisfies 
the condition $\xi_{a}\oder{x^a}{\l}=0$. 
Because $\xi^a$ is a Killing vector field on the spacetime $(M,fg)$,
$\xi_{a}\oder{x^a}{\l}$ is a conserved quantity. 
Therefore, the condition 
$\xi_{a}\oder{x^a}{\l}=0$ is consistent with the dynamics.

\begin{table*}[t]
\caption{
  Integrability 
  of C1 strings on $\ads_5$. 
  Any 
  {\hv} 
  $\xi$ is equivalent 
  to one of ten types~\cite{Koike:2008fs}. 
  For each type, there is 
  a three-dimensional 
  Abelian Lie subalgebra $\CC$ 
  containing $\xi$, which is shown by its generators
  $\angl{\bullet,\bullet,\bullet}$. 
  Each type has a hidden conserved quantity 
  $Q:=K^{ab}P_a P_b+K_0$. 
  The set of five mutually commuting conserved quantities, 
  $\normalparen{g^{ab}P_aP_b,\;
    \xi^aP_a,\;
    X\sbp 1^aP_a,\;
    X\sbp 2^aP_a,\;
    Q}$,  
  implies integrability of each C1 string. 
  Notation: 
  $(x^i)
  =(s,t,x,y,z,w)$ and 
  $L$, $J_{xy}$, $\wt K_z$, $K_w$ 
  are generators of 
  the $st$ rotation, 
  the $xy$ rotation, 
  the $sz$ boost, 
  the $tw$ boost, etc.  
  The product $AB$ 
  denotes
  the symmetric tensor product, 
  $(AB)^{ab}:=(A^aB^b+B^aA^b)/2$. 
} 
\begin{center}
\begin{tabular}
{lp{5.7cm}cp{6.7cm}cp{2.9cm}} 
\hline \hline
type & $\xi$ and Abelian subalgebra $\CC\ni\xi$& & $K^{ab}$ & & $K_0$
\\ \hline
$(4|0)$ &
$\Lxs+\Lyz+\Lzs+\Lst$; \qq\qq\qq\qq\qq\linebreak
$\bigl\langle
\Lxs+\Lyz+\Lzs+\Lst$, 
$\Lxy+\Lyz-\Lyt+\Lzs$, 
$\Lxw-\Lwt 
\bigr\rangle$
& &
$
2(\Lxz-\Lzt)(\Lxt+\Lys)
+2\Lzw(\Lxw-\Lwt)    \qq\qq\linebreak
+2(\Lxy-\Lyt)^2+(\Lxz-\Lzt)^2+(\Lyz+\Lzs)^2 \linebreak
-(\Lyw+\Lws)^2-(\Lxy-\Lxs)^2+(\Lyt+\Lst)^2$ 
& &
$(\Xt-\Xx)^2$
\\ \hline
$\pm(3,1|0)$ &
$\Lxt+\Lys+\Lyz\pm\Lxw  
+a(\Lxy-\Lst\mp\Lzw)$;  \linebreak
$\bigl\langle
\Lxt+\Lys+\Lyz\pm\Lxw,  \qq\qq\qq\qq\qq\linebreak
\Lxy-\Lst\mp\Lzw,  
\mp\Lzw-\Lzt\pm\Lws+\Lst
\bigr\rangle$
& &
$(\Lxz+\Lxs\pm\Lyw+\Lyt)^2  
+(\pm\Lxw+\Lxt-\Lyz-\Lys)^2   \qq\linebreak
+4a[(\Lxz+\Lxs\pm\Lyw+\Lyt)(\Lzs\mp\Lwt) \qq\qq\linebreak
{}\qq
+(\pm\Lxw+\Lxt-\Lyz-\Lys)(\Lzt\pm\Lws)]
$
& &
$-4a^2
\qq\qq\linebreak
\times[(\Xz-\Xs)^2+(\Xw\mp\Xt)^2]$
\\ \hline
$(2,2|0)$ &   
$\Lxt+\Lst+a\Lyz$; \qq\qq\qq\qq\qq\qq \linebreak
$\bigl\langle
\Lxt+\Lst, 
\Lyz, 
\Lxw+\Lws
\bigr\rangle$
& &
$
(\Lxy+\Lys)^2+(\Lxz+\Lzs)^2   \qq\qq\qq\qq\qq\linebreak
+a^2(\Lxw^2-\Lxs^2-\Lxt^2-\Lws^2-\Lwt^2+\Lst^2)
$
& &
$a^2(\Xx+\Xs)^2$
\\ \hline
$(2,-2|0)$ &
$\Lxt+\Lxy+a\Lzw$;   \qq\qq\qq\qq\qq\qq \linebreak
$\bigl\langle
\Lxt+\Lxy$, 
$\Lzw$, 
$\Lys+\Lst
\bigr\rangle$
& &
$
-(\Lyz-\Lzt)^2-(\Lyw-\Lwt)^2   \qq\qq\qq\qq\qq\linebreak
+a^2(\Lxy^2-\Lxs^2-\Lxt^2-\Lys^2-\Lyt^2+\Lst^2)
$
& &
$-a^2(\Xy-\Xt)^2$
\\ \hline
$(0|2)$ &
$\Lxt+\Lxy+a\Lzs$; \qq\qq\qq\qq\qq\qq \linebreak
$\bigl\langle
\Lxt+\Lxy$, 
$\Lzs$, 
$\Lyw-\Lwt
\bigr\rangle$
& &
$
 -(\Lyz-\Lzt)^2+(\Lys+\Lst)^2   \qq\qq\qq\qq\qq\linebreak
 +a^2(-\Lxy^2-\Lxw^2+\Lxt^2-\Lyw^2+\Lyt^2+\Lwt^2)
$
& &
$a^2(\Xy-\Xt)^2$
\\ \hline
$(2,1,1|0)$ & 
$\Lxt+\Lys+\Lxy+\Lst 
+a\Lzw      
+b(\Lxy-\Lst)$;    \linebreak
$\bigl\langle
\Lxt +\Lys+\Lxy+\Lst, 
\Lzw,
\Lxy-\Lst
\bigr\rangle$
& &
$(a^2-b^2)[(\Lxs+\Lyt)^2+(\Lxt-\Lys)^2]   \qq\qq\qq\qq\linebreak
 +2b
[(\Lxz+\Lzs)^2+(\Lyz-\Lzt)^2 \qq\qq\qq\qq \linebreak
{}\qq
 +(\Lxw+\Lws)^2
 +(\Lyw-\Lwt)^2]
$
& &
$2b(a^2-b^2)   \qq\qq\qq\linebreak
\times [(\Xx+\Xs)^2   
+(\Xy-\Xt)^2]$
\\ \hline
$(2|1)$ &
$
\Lxt+\Lys+\Lxy+\Lst 
+a\Lzw       
+b(\Lyt+\Lxs)
$;          
\linebreak
$\bigl\langle
\Lxt+\Lys+\Lxy+\Lst, 
\Lzw,
\Lyt+\Lxs
\bigr\rangle$
& &
$ (a^2+b^2)[(\Lxy-\Lst)^2-(\Lxt-\Lys)^2]   
 +4b 
\qq\qq\qq\linebreak
\times
[(\Lxz+\Lzs)(\Lyz-\Lzt)  
 +(\Lxw+\Lws)(\Lyw-\Lwt)]
$ & &
$4b(a^2+b^2)
\qq\qq\qq\linebreak
\times(\Xx+\Xs)(\Xy-\Xt)$
\\ \hline
$(1,1,1,1|0)$ & 
$a\Lst+b\Lxy+c\Lzw$; \qq\qq\qq\qq\qq\linebreak
$\bigl\langle
\Lst,\Lxy,\Lzw
\bigr\rangle$
& &
$
(a^2-b^2)(\Lxz^2+\Lxw^2+\Lyz^2+\Lyw^2)  \qq\qq\qq\qq\linebreak
 +(b^2-c^2)(\Lxs^2+\Lxt^2+\Lys^2+\Lyt^2)
$
 & &
$-(a^2-b^2)(b^2-c^2)  
\linebreak
\times(\Xx^2+\Xy^2)$
\\ \hline
$(1,1|1)$ &
$
\Lxt+\Lys 
+b(\Lst-\Lxy)+c\Lzw
$;   \qq\qq\qq\linebreak
$\bigl\langle
\Lxt+\Lys$, 
$\Lst-\Lxy$,
$\Lzw
\bigr\rangle$
 & &
$
(b^2-c^2-1)[(\Lxs+\Lyt)^2+(\Lxt-\Lys)^2]   \qq\qq\qq\linebreak
 +4b(\Lxz\Lzs+\Lxw\Lws-\Lyz\Lzt-\Lyw\Lwt)
$
& &
$-4b[(b^2-c^2-1)   \qq\qq\linebreak 
(\Xx\Xs-\Xy\Xt) 
-b(\Xz^2+\Xw^2)]$
\\ \hline
$(0|1,1)$ &
$a\Lxt+b\Lys+c\Lzw$;   \qq\qq\qq\qq\qq\linebreak
$\bigl\langle
\Lxt,\Lys,\Lzw
\bigr\rangle$
& &
$
 (b^2+c^2)(\Lxz^2+\Lxw^2-\Lzt^2-\Lwt^2)   \qq\qq\qq\qq\linebreak
 +(a^2+c^2)(\Lyz^2+\Lyw^2-\Lzs^2-\Lws^2)
$
 & &
$-(a^2+c^2)(b^2+c^2)  
\linebreak 
\times (\Xz^2+\Xw^2)$
\\ \hline \hline
\end{tabular}
\end{center}
\label{tab-ads5}
\end{table*}

\se{C1-string integrability}
Let us move 
to the Hamiltonian formalism, which is suitable 
for systematic discussions of the conserved quantities. 
The Hamiltonian for 
the action \eqref{eq-lifted-action} 
is given by 
\begin{align}
  H=\f l{4\pi\a'}N\paren{\f1f g^{ab}P_aP_b+1}. 
  \label{eq-H}
\end{align}
The geodesics on $(O,fh)$ is given by the solutions of 
the Hamiltonian system with 
the lifted Hamiltonian \Ref{eq-H} that satisfy the condition 
$\xi^aP_a=0$. 

The Hamiltonian system 
\eqref{eq-H} with 
$2d$-dimensional phase space 
is said to be \emph{Liouville integrable} 
if there exist $d$ independent 
conserved quantities $Q_i$, $0\le i\le d$, 
such that 
they Poisson-commute with each other. 
If the system is Liouville integrable, 
the equation of motion can be solved by quadrature. 
We will say that the spacetime $(M,g)$ is \emph{C1-string integrable} 
if all C1 strings on $(M,g)$ are Liouville integrable. 

Let us 
recast 
the geodesic Hamiltonian \eqref{eq-H} into the 
``potential form,'' by 
choosing $N\equiv f$: 
\begin{align}
  \wt H:= 
  \f l{4\pi\a'}
  \paren{ g^{ab}P_aP_b+f }, 
  \q
  \xi^aP_a=0. 
  \label{eq-H-pot}
\end{align}
The advantage of the potential form is that the coefficients  
in the kinetic term recover the original 
metric $g$. 
This enables one 
to take the full advantage of the symmetry of $(M,g)$. 
A conserved quantity $Q$ which is 
a polynomial in $P_a$ 
can be obtained by 
the \emph{Killing hierarchy}~\cite{K_Hierarchy}, 
which is obtained by comparing the coefficients 
of each order of $P_a$ in the equation 
$\Pb {\wt H}Q=0$, 
where $\Pb\bullet\bullet$ denotes the Poisson bracket. 
The simplest of such conserved quantities is 
$Q:=X^aP_a$ which is first order in $P_a$. 
For such $Q$ and the Hamiltonian 
\Ref{eq-H-pot}, 
the Killing hierarchy reads
\begin{align}
  \nabla^{(a}X^{b)}=0, \q 
  X^a\nabla_af=0. 
  \label{eq-KH}
\end{align}
The first equation implies that $X^a$ is a Killing vector on
$(M,g)$, 
while 
the second 
implies that the norm of $\xi$ does not change in the
direction of $X$. 

Now, let us consider 
the case where $Q$ is a second order polynomial 
in $P_a$. 
One observes that the odd- and even-order equations 
in the Killing hierarchy do not couple 
because $\wt H$ in \Ref{eq-H-pot} 
has even-order terms only. 
Thus it is sufficient to 
consider $Q:=K^{ab}P_aP_b+K_0$. 
The condition 
$\normalbrac{\wt H,Q}=0$  
is equivalent to 
\begin{align}
  \nabla^{(a} K^{bc)}=0,\q
  \nabla^{a} K_0&=K^{ab}\nabla_b f. 
  \label{eq-kh-2o}
\end{align}
The first equation states that 
$K^{ab}$ is a \emph{Killing tensor field}. 
Given a Killing tensor $K^{ab}$, a function 
$K_0$ that satisfies the latter 
exists if and only if the consistency condition 
\begin{align}
  \nabla^{[a} \bigparen{K^{b]c}\nabla_c f}
  =0 
  \label{eq-consistency}
\end{align}
is satisfied~\cite{K_Hierarchy}. 
The quantities 
$X^aP_a$ and $Y^aP_a$ Poisson-commute if 
the vector fields $X$ and $Y$ commute. 
The quantities $X^aP_a$ and $K^{ab}P_aP_b+K_0$ 
Poisson-commute 
if 
$\LL_X K^{ab}=0$ and $X^a\nabla_aK_0=0$. 

As stated above, 
a C1 string is Liouville integrable if 
there exist $d$ mutually commuting conserved quantities for $\wt H$ in
\Ref{eq-H-pot}. 
The Hamiltonian $\wt H$ 
has at least two commuting 
conserved quantities, $\wt H$ itself and $\xi^aP_a$. 
We therefore 
propose the following procedure to examine the Liouville
integrability of a C1 string. 

Step1:  
  Enumerate independent 
  Killing vectors 
  $X\sbp i$, $1\le i\le I$, 
  on $(M,g)$ 
  which 
  commute 
  with $\xi$ 
  and 
  with each other. 
  Then 
  $\normalparen{\wt H, 
    \xi^aP_a, X\sbp 1^aP_a,..., 
    X\sbp I^aP_a}
  $ 
  is a set of conserved quantities 
  that Poisson-commutes with each other. 
  If $2+I\ge d$, the C1 string is Liouville integrable. 

Step 2: 
  If $2+I<d$, seek for second order conserved quantities 
  $Q\sbp j=K^{ab}\sbp jP_aP_b+K_0{\sbp j}$, $1\le j\le J$, 
  where $K^{ab}\sbp j$ is a Killing tensor 
  on $(M,g)$. 
  If $K^{ab}$ satisfies the consistency condition 
  \Ref{eq-consistency}, 
  there is $K_0$ such that $Q$ 
  is conserved. Check that $Q\sbp j$ commute 
  with each other and with the other conserved quantities. 
  If $2+I+J\ge2$, the C1 string is Liouville integrable. 

If the spacetime $(M,g)$ is maximally symmetric, the 
Killing tensor $K^{ab}$ in the second step is always \emph{reducible}, 
i.e., it is a linear combination of symmetric tensor products of
Killing vectors. 
We remark that one can in principle 
continue the procedure above for
higher order conserved quantities.

\se{C1 surfaces in $S^5$}
\label{s5}
The first space that we 
want 
to 
examine 
C1-string integrability is the 5-dimensional sphere $S^5$. 
The space $S^5$ is a hypersurface
$\sum_{i=1}^{6}(x^i)^2=1$ 
in six-dimensional flat space
with Cartesian coordinates $(x^i)$, $1\le i\le 6$. 
The isometry group $G$ is $O(6)$. 
Let us first classify 
the {\hv}s $\xi$ up to conjugacy of $G$. 
In the case of $S^5$, 
there is only one type of them, 
\begin{align}
  \xi=
  a_1 L_{12}
  +
  a_2 L_{34}
  +
  a_3 L_{56}, 
  \label{eq-s5-xi}
\end{align}
where each 
$L_{ij}$ denotes the generator of the rotation in the 
$x^i x^j$ plane and $a_1$, $a_2$ and $a_3$ are real numbers. 
This may be most easily seen by the fact that 
$\alg o(6)$ is isomorphic to $\alg s\alg u(4)$.
Since any {\sa} 
matrix can be 
diagonalized by a unitary matrix, 
$\alg s\alg u(4)/\Ad_{\mathrm {SU(4)}}$ can be identified with the 
set of traceless real diagonal matrices, 
which is a Cartan subalgebra of $\alg o(6)$. 
A basis 
thereof 
correspond to 
$\brac{ L_{12},L_{34},L_{56} }$ appearing in \Ref{eq-s5-xi}. 

From this Cartan subalgebra, 
one 
can choose a set 
$(\xi, X\sbp 1, X\sbp 2)$ 
of 
linearly independent, mutually commuting 
Killing vectors. 
Then we have a set 
$(\wt H, 
\xi^aP_a, 
X\sbp 1^aP_a, 
X\sbp 2^aP_a)$ 
of 
four mutually commuting conserved quantities. 
One more conserved quantity 
is necessary for integrability, so that we go into Step 2
in the procedure above. 
Since $S^5$ is maximally symmetric, any Killing tensor is reducible. 
We observe that 
$M_{ij}:=L_{{2i-1},{2j-1}}^2+L_{{2i-1},{2j}}^2
+L_{{2i},{2j-1}}^2+L_{{2i},{2j}}^2$ 
($1\le i<j\le 3$) commute with 
$L_{12}$
$L_{34}$ and
$L_{56}$. 
Let us therefore try the reducible Killing tensor 
\begin{align}
  K^{ab} 
  & 
  = 
  c_{12}M_{12}
  +c_{13}M_{13}
  +c_{23}M_{23}, 
\end{align}
where $c_{12},c_{13},c_{23}$ are real numbers. 
The consistency condition \Ref{eq-consistency} holds 
if the coefficients $c_{ij}$ satisfy 
\begin{align}
  (a_1^2-a_2^2)c_{12}+(a_3^2-a_1^2)c_{13}+(a_2^2-a_3^2)c_{23} = 0. 
  \label{eq-s5-consis-coef}
\end{align}
Since such $c_{ij}$ always exist, 
there is a conserved quantity 
$Q=K^{ab}P_aP_b+K_0$ which commutes with the 
others~\cite{fn-s5}. 
Therefore, 
$S^5$ is 
C1-string integrable, namely, 
all C1 minimal surfaces can be obtained by quadrature.

\se{C1 strings in $\ads_5$}
\label{ads}
The five-dimensional 
anti-de Sitter spacetime 
$\ads_5$ is a pseudosphere 
$\eta_{ij} x^i x^j=-1$, where 
$\eta_{ij}=\diag[-1,-1,1,1,1,1]$, 
in the space $(x^i)_{1\le i\le 6}$
$=(s,t,x,y,z,w)$ 
with the flat metric $ds^2=\eta_{ij}dx^idx^j$. 
The isometry group $G$ is $O(4,2)$. 
In contrast to the case of $S^5$, 
the {\hv} $\xi$ has a rich variety. 
They are classified into ten types~\cite{Koike:2008fs} 
which are shown in Table~\ref{tab-ads5}, 
where 
$J_{x^{i}x^{j}}:=x^i \partial_{x^{j}} - x^j \partial_{x^{i}}$, 
$K_{x^{i}}:=t\partial_{x^{i}}+x^{i}\partial_{t}$ 
$\widetilde{K}_{x^{i}}:=s\partial_{x^{i}}+x^{i}\partial_{s}$, 
and $L:=s\partial_t-t\partial_s$
denote 
rotation, $t$-boost, $s$-boost, and $st$-rotation,
respectively.
For each type of $\xi$, 
the situation concerning to integrability 
is similar to the case of 
$S^5$. 
We can choose a set 
$(\xi, X\sbp 1, X\sbp 2)$ 
of 
linearly independent, mutually commuting 
Killing vectors 
so that we have a set 
$(\wt H, 
\xi^aP_a, 
X\sbp 1^aP_a, 
X\sbp 2^aP_a)$
of four mutually commuting conserved quantities. 
One conserved quantity is missing for integrability, 
but there is a second order conserved quantity which commutes with
them. 
We carried out Step 2 for each type of $\xi$ 
and found $K^{ab}$ that satisfies the consistency
condition \Ref{eq-consistency}. 
Table~\ref{tab-ads5} shows 
the resulting conserved quantity 
$Q=K^{ab}P_aP_b+K_0$ for each class. 
This proves the C1-string integrability of the spacetime $\ads_5$.

\se{
  C1 strings on $\ads_5\times X^5$} 
\label{direct}
Let us consider a general product spacetime 
$(M,g)
=(M_1,g_1)\times (M_2,g_2)
=(M_1\times M_2, g_1\oplus g_2)
$

whose  
isometry group 
can be written as 
$
G=G_1\times G_2
$, where 
each 
$G_i$ is the isometry group of $(M_i,g_i)$. 
Then any C1 string on $M$ 
is defined by a Killing vector field 
$\xi=\xi_1+\xi_2$, where $\xi_i\in\alg g_i$, 
where $\alg g_i$ is the Lie algebra of $G_i$. 
Thus we have 
\begin{align}
  f=f_1+f_2, 
\end{align}
where $f_i:=\norm{ \xi_i }_{g_i}^2$. 
This means that the potential in the Hamiltonian 
$\wt H$ given by \Ref{eq-H-pot} splits. 
Therefore, 
the Hamiltonian $\wt H$ 
admits a separation of variables, 
$
  \wt H=\wt H_1 +\wt H_2, 
$
where 
each 
$\wt H_i$ is a Hamiltonian of the potential form \Ref{eq-H-pot}
on the space $(M_i,g_i)$. 
We conclude that 
a C1 string defined by $\xi$ is Liouville integrable if and only if  
the C1 strings 
defined by 
$\xi_i$ are Liouville integrable.

The spacetime with 
$M_1=\ads_5$ and $M_2=X^5$, 
where $X^5$ is a
five-dimensional Riemannian manifold, 
is of special interest 
in recent studies in string
theory and gravity. 
The simplest such case is $X^5=S^5$, 
where 
$G_1=\Orth(4,2)$ and $G_2=\Orth(6)$. 
Because 
both $\ads_5$ and $S^5$ are C1-string integrable, 
the argument above immediately implies that 
$\ads_5\times S^5$ are 
C1-string integrable.

Another important case is $X^5=T^{p,q}$. 
The metric on $T^{p,q}$ is given by 
\begin{align}
  g=
  &
  \sum_{i=1,2}\a_i^2(d\tht_i^2+\sin^2\tht_id\phi_i^2)
  \nn
  &+
  \b^2(d\psi+q_1\cos\tht_1d\phi_1+q_2\cos\tht_2d\phi_2)^2, 
  \label{eq-tpq-g}
\end{align}
where $q_1:=p$, $q_2:=q$. 
By the symmetry of 
exchanging the indices $i=1$ and $i=2$, 
we assume $p\le q$ without loss of generality. 

The group $G_2$ is seven-dimensional and given by 
$
  G_2=\SO(3)\times\SO(3)\times \U(1) 
$
up to discrete isometry
whose generators 
are 
\begin{align}
  &L\spp i_{x}
  :=
  \sin\phi_i \p_{\tht_i}
  +
  \f{\cos\phi_i}{\tan\tht_i}
  \paren{\p_{\phi_i}-\f{q_i\p_\psi}{\cos\tht_i}}, 
  \\
  &L\spp i_{y}
  :=
  \cos\phi_i \p_{\tht_i}
  -
  \f{\sin\phi_i}{\tan\tht_i}
  \paren{\p_{\phi_i}-\f{q_i\p_\psi}{\cos\tht_i}}, 
  \\
  &L\spp i_{z}
  :=\p_{\phi_i}, 
  \\
  &L_{\psi}
  :=\p_\psi, 
\end{align}
where $i=1,2$, 
except for a special case in $T^{0,1}$ 
in which case the isometry
group becomes larger~\cite{fn-enh}. 
Any {\hv} can be written as 
\begin{align}
  \xi=a_1 L\spp 1_{z}+a_2 L\spp 2_{z}+b \, L_{\psi}  
  \label{eq-kv-tpq}
\end{align}
up to conjugacy of $G_2$. 

Before analyzing the C1 strings on 
$\ads_5\times T^{p,q}$, 
we recall that motion of a free particle on 
the spacetime 
is integrable. 
First, the spacetime $\ads_5$ is geodesically integrable. 
Second , the space $T^{p,q}$ 
has three commuting Killing vectors 
$L\spp 1_{z}$, $L\spp 2_{z}$, and $L_{\psi}$. 
The Casimir operator of the first factor $\SO(3)$ 
defines a reducible Killing tensor $K^{ab}$ which commutes with
all of them. 
Then the quintet 
$\normalparen{g^{ab}P_aP_b,\;L\spp 1_z{}^aP_a,\;L\spp 2_z{}^aP_a, 
L_{\psi}^{a}P_{a},\;K^{ab}P_aP_b}$ consists of 
mutually commuting, independent 
conserved quantities. 
Thus $T^{p,q}$ is geodesically integrable. 
Therefore, spacetimes $\ads_5\times T^{p,q}$ 
are also geodesically integrable. 
By contrast, 
the behavior of the strings differs. 
There is a 
C1 
string that is not integrable on 
$\ads_5\times T^{1,1}$~\cite{Basu:2011di
}. 
We 
examine 
C1-string integrability of 
each $\ads_5\times T^{p,q}$ and 
see that C1-string integrability 
depends on $(p,q)$. 

The Hamiltonian $\wt H$ on $T^{p,q}$ has 
the potential
\begin{align}
  f=
  &
  \sum_{i=1,2}
  \paren{
    a_i^2\paren{ \alpha_i^2 \sin^2\tht_i+q_i^2 \beta^2 \cos^2\tht_i }
    +
    2q_ia_ib \beta^2 \cos\theta_i
  }
  \nn
  &
  +b^2 \beta^2
  +
  2a_1a_2q_1q_2 \beta^2 \cos\theta_1\cos\theta_2. 
  \label{eq-Tpq-pot}
\end{align}
The Killing vectors 
$L\spp1_z$, 
$L\spp2_z$ and 
$L_{\psi}$ 
appearing in 
\Ref{eq-kv-tpq} 
are mutually commuting 
so that 
we can construct, by 
their linear combinations, a set 
$(\xi, X\sbp 1, X\sbp 2)$ of 
three linearly independent, mutually commuting 
Killing vectors. 
Thus we have a set of 
four mutually commuting conserved quantities 
$(\wt H, 
\xi^aP_a, 
X\sbp 1^aP_a, 
X\sbp 2^aP_a)$. 
We seek for one more conserved quantity, which is second 
order in $P_a$ (Step 2). 
There is no irreducible Killing tensor on $T^{p,q}$ which commutes 
with $\xi$. 
Reducible Killing tensors which commute with $\xi$ are
\begin{align}
  K\sbp i=&
  L\psp i_{x}{}^2+
  L\psp i_{y}{}^2+
  L\psp i_{z}{}^2
  -q_i^2 L_{\psi}^2
  \nn 
  =&
  \p_{\tht_i}^2
  +\f1{\sin^2\tht_i}
  \paren{\p_{\phi_i}-q_i \cos\tht_i \p_\psi}^2, 
  \q i=1,2. 
\end{align}
They are related to the (inverse) metric by 
$g^{ab}=K^{ab}\sbp1/{\a_1^2}+K^{ab}\sbp2/{\a_2^2}
+L_{\psi}^{a}L_{\psi}^{b}/{\b^2}$.  
Thus $K\sbp1$ and $K\sbp2$ are not independent and it is enough to
examine if there is a conserved quantity $Q=K\sbp1^{ab}P_aP_b+K_0$  
(rather than to consider the Killing hierarchy for all combinations of
$K\sbp1$ and $K\sbp2$). 
We observe that  
the potential 
$f$ 
becomes separable, 
$
  f=f\sbp1(\theta_1)+f\sbp2(\theta_2), 
$
for any {\hv} $\xi$, 
if the last term vanishes in the expression 
\Ref{eq-Tpq-pot}. 
In that case, 
it follows immediately that 
$K\sb0=(\a_1^2/a_1^2)f\sbp1$ is a solution to 
the latter equation in the Killing hierarchy \Ref{eq-kh-2o} 
with $K^{ab}=K\sbp1^{ab}$. 
Thus the quintet 
$(\wt H, 
\xi^aP_a, 
X\sbp 1^aP_a, 
X\sbp 2^aP_a, Q)$ is a set of mutually commuting conserved
quantities and proves the integrability of the C1 string defined by
$\xi$. 
On the other hand, 
if the last term 
in the expression 
\Ref{eq-Tpq-pot} 
does not vanish, 
the C1 string 
defined by
$\xi$ is not integrable. 
The potential 
$f$ including the term which depends on both $\tht_1$ and $\tht_2$
fails to satisfy 
the consistency condition \Ref{eq-consistency} 
with $K^{ab}=K\sbp1^{ab}$~\cite{fn-consis-no}. 
To summarize, 
the C1 string 
defined by
$\xi$ is integrable if and only if 
the last term in \Ref{eq-Tpq-pot} is identically zero. 

This leads to the following results. 
First, the space $T^{0,q}$ is C1-string integrable. 
Thus, by the argument above on direct products, 
$\ads_5\times T^{0,q}$ is C1-string integrable. 
Second, 
the space $T^{p,q}$ with $pq\ne0$ is not C1-string integrable 
(as far as conserved quantities in order two or less in 
$P_a$ are concerned). 
Namely, the C1 strings defined 
by $\xi=a_1L_{z}\spp1+bL_{\psi}$ and 
by $\xi=a_2L_{z}\spp2+bL_{\psi}$ 
are integrable 
but the other C1 strings 
are not integrable. 
Thus, 
$\ads_5\times T^{p,q}$ ($pq\ne0$) is not C1-string integrable. 
We remark that the metric \Ref{eq-tpq-g} on $T^{p,q}$ 
satisfies the property called Sasaki-Einstein
and $\ads_5\times T^{p,q}$ partially retains supersymmetry
when $p=q=1$,   
$\a_1^2=\a_2^2=1/6$, and $\b^2=1/9$. 
C1 integrability does not depend on whether the
spacetime retains supersymmetry or not.

\se{Conclusion} 
We presented a method with which one can systematically 
analyze conserved quantities and integrability of strings of class C1
(cohomogeneity-one). 
We applied the method to several important spaces and spacetimes,  
$S^5$, $\ads_5$, $\ads_5\times S^5$, and $\ads_5\times T^{p,q}$. 
We found that the first three space and spacetimes are C1-string
integrable, in which 
\emph{all} C1 strings are Liouville integrable. 
On the other hand, 
C1-string integrability of $\ads_5\times T^{p,q}$, 
depends on $(p,q)$. 
Namely, $T^{0,q}$ are C1-string integrable while 
$T^{p,q}$ with $pq\ne0$ are not. 
Thus the 
chaotic behavior in $\ads_5\times T^{1,1}$ is 
naturally understood as absence of 
necessary conserved quantity for C1-string nonintegrability.

As in the example of Kerr spacetime, 
geodesic integrability 
can probe hidden symmetry 
which is not implied by the isometry group. 
C1-string integrability discriminates 
$\ads_5\times T^{p,q}$, 
which 
have the same isometry group 
and are geodesically integrable. 
Therefore, 
C1-string integrability (or nonintegrability) probes finer 
hidden symmetry. 
Intuitively, 
by the C1-string integrability, 
one is measuring the remaining symmetry after a ``piece'' of
symmetry (corresponding to the {\hv}) is removed. 
One can generalize the C1-string integrability 
to higher-dimensional objects, C1-membrane integrability and so forth, 
which might bring further useful information of the spacetime.

The authors thank Y.~Yasui and K.~Yoshida for useful discussions. 
This work was supported by JSPS KAKENHI Grant Number JP16K05358 (HI).


\begin{thebibliography}{99}
\bibitem{fn-s5} 
{  An example is 
  $c_{12}=(a_1^2-a_3^2)$, $c_{13}=(a_1^2-a_2^2)$, $c_{23}=0$ 
  which leads to 
  $
  K_0 = (a_1^2-a_2^2) (a_1^2-a_3^2) (x_1^2+x_2^2)
  $
  hence to the corresponding $Q$. 
  Other choices of $c_{ij}$ give conserved quantities 
  which are constructed by $Q$ and 
  $
  (
  \tilde{H}, 
  \xi^aP_a, 
  X\sbp 1^aP_a, 
  X\sbp 2^aP_a)$
}.
\bibitem{fn-enh} 
{  In the case $T^{0,1}=S^2\times S^3$ (topologically) 
  and $\a_2=\b$, 
  the symmetry enhances 
  to $G_2=\SO(3)\times \SO(4)$. 
  Although 
  there might be in principle 
  a {\hv} $\xi$ 
  different from \Ref{eq-kv-tpq}, 
  actually $\xi$ in this case is always given by \Ref{eq-kv-tpq} 
  up to conjugacy of the enhanced $G_2$. 
}
\bibitem{fn-consis-no}
{Eq.~\Ref{eq-consistency} states that 
$\om_b:=K_{b}{}^c\nabla_c f$ is a closed 1-form}. 
However, 
in coordinates $(\tht_1,\phi_1,\tht_2,\phi_2,\psi)$, 
one has $\om_{\tht_2}=0$ while 
$\om_{\tht_1}$ depends on $\tht_2$. 
Thus 
$\om_b$ cannot be closed.
\end{thebibliography}
\end{document}